\documentclass[12pt]{article}
\newcommand{\T}{\cal{T}}
\begin{document}
\begin{center}
{\large The Role of Energy and a New Approach} \\
{\large to Gravitational Waves in General Relativity}
\\
F. I. Cooperstock
 \\{\small \it Department of Physics and Astronomy, University 
of Victoria} \\
{\small \it P.O. Box 3055, Victoria, B.C. V8W 3P6 (Canada)}\\
{\small \it Permanent e-mail address: cooperstock@phys.uvic.ca }
\end{center}
\begin{abstract}
The energy localization hypothesis of the author that energy is localized in non-vanishing regions of the energy-momentum tensor implies that gravitational waves do not carry energy in vacuum. If substantiated, this has significant implications for current research. Support for the hypothesis is provided by a re-examination of Eddington's classic calculation of energy loss by a spinning rod. It is emphasized that Eddington did not monitor the entire Tolman energy integral, concentrating solely upon the change of the `kinetic' part of the energy. The `quadrupole formula' is thus seen to measure the kinetic energy change. When the derivative of the missing stress-trace integral is computed, it is seen to cancel the Eddington term and hence the energy of the rod is conserved, in support of the localization hypothesis. The issue of initial and final states is addressed.
\end{abstract}
 
.

\mbox{ }\\
{\large 1. Introduction}

      From the time of its inception more than 80 years ago, general
relativity has presented problems in connection with the energy concept. In 
other areas of physics, there were tensorial conservation laws both locally and 
globally. However, in general relativity, global energy-momentum conservation 
was accessible only through the aid of pseudotensors with their attendant 
ambiguities. Many researchers from the time of Einstein who introduced them, 
were willing to deal with these less-than-satisfactory constructs, albeit with 
an underlying sense of unease. While they did present a certain degree of 
consistency when used in a particular manner (with asymptotic flatness and
Cartesian coordinates), the fact that they could be annihilated at will at any 
given space-time point by the right choice of coordinates, rendered the issue of 
energy localization in general relativity problematic. Einstein \cite{einst}and Eddington 
\cite{edd} accepted the apparently inherent non-localizability of gravitational energy on 
the basis of the nature of the pseudotensor. Through the years, various authors 
\cite{motion} chose to concentrate their efforts on direct analyses of the 
motion of bodies in general relativity, motivated at least in part by the 
inherent ambiguities presented by the energy concept in general relativity.   

However, the energy concept is fundamental to physics and the challenge to 
understand it more clearly within the context of general relativity continued to 
attract the attention and effort of various researchers \cite{energyauth}.
Interestingly, Bondi, one of the prominent pioneers in the field, recently returned to the question, arguing in favour of the necessity for energy localization \cite{bondinew}. Apart 
from the fundamental importance of the issue, there were practical 
considerations as well. The elusive gravitational waves, first proposed by 
Einstein to be emitted by accelerated masses in analogy with electromagnetic 
waves from accelerated charges in electromagnetism, had escaped all but a 
conjectured indirect indication of their presence from the period variation of 
the binary pulsar. 
If there were to be a direct detection, the fundamental nature of these waves 
vis-\`{a}-vis energy content would have to be understood.

In recent years, various factors led this author to a new hypothesis for the 
localization of energy in general relativity \cite{coop}. The generally 
covariant energy-momentum conservation laws in general relativity
which are devoid of content in vacuum, acquire apparent content in vacuum 
when re-expressed as an ordinary divergence with the pseudotensor and
the Gauss theorem employed. Apparent wave energy is computed by a flux emanating 
from the pseudotensor in the asymptotic region yet the originating 
equation is 
actually devoid of content in vacuum. This suggests that the pseudotensor might 
be
injecting ``pseudo-content", which would not be surprising, given its ephemeral nature.
 Plane gravitational waves are traditionally expressed in the ``transverse-
traceless" gauge in which the pseudotensor reveals an apparent energy flux. 
However, since such waves are in the Kerr-Schild class, they can be expressed in 
a 
form in which all components of their pseudotensor vanish globally 
\cite{gur}. Then 
they indicate a total lack of energy or momentum, unlike their electromagnetic 
plane-wave cousins which have a tensorially covariant energy and momentum 
content. Particle physicists generally tend to view gravity as just another field and the graviton as just another zero-rest-mass particle. However, it is to be emphasized that all particles and fields
apart from gravity exist \textit{within} spacetime whereas, in essence, gravity \textit{is} spacetime. From the general relativity perspective, gravity assumes a very special role.
These and other facts \cite{coop} led the author to hypothesize that \textit{in 
generality, energy and 
momentum are localized in regions of the energy-momentum tensor $T_i^k$}. This 
would imply that \textit{gravitational waves are not carriers of energy and 
momentum in 
vacuum} \cite{coop}.  If correct, this would have far-reaching consequences. Apart from a revamping of basic concepts in classical general relativity, 
the nature of the quantization of gravity would have to be re-evaluated: for what is a graviton in the absence of energy?
Various earlier demonstrations to the contrary of the hypothesis were 
analyzed and their flaws indicated \cite{coopand}
but it was evident that a clear-cut demonstration in favour of the hypothesis 
was called for. This is the primary aim of the present paper.

To this end, we focus upon one of the very earliest calculations of supposed 
energy loss by a 
dynamical system in the form of a spinning rod \cite{edd}. We first show that 
the complete mass function was not monitored by this calculation and that a 
vital missing part \cite{tol} is required to do so. We 
achieve this by a modification of a method devised by Einstein. A remarkable 
thing emerges when the missing Tolman terms are accounted for: \textit{the 
supposed energy loss first calculated by Eddington is precisely cancelled by the 
missing Tolman terms in support of our localization hypothesis}.  
While we analyzed this particular system because of its direct physical clarity, the result can be generalized in the manner of \cite{perron}.

In sec. 2, we build the mathematical framework for the theory. In sec. 3, we 
review the ingenious calculation of Eddington for the spinning rod and in the process, 
indicate where he had made an unjustified assumption in pre-supposing the 
negligible character of certain terms. In actuality, these terms 
which Eddington neglected undergo a fortuitous cancellation. In sec. 4, we 
review the achievement of Tolman in expressing the energy of a static or quasi-static 
system as an integral over the region of the energy-momentum tensor. In sec. 5, we use 
a modification of the method of Einstein to express the contributions to mass 
which had been neglected previously. These are seen to cancel the original 
Eddington portion of the mass-loss function. A summary and concluding discussion are presented in sec. 6.\\
 \mbox{ }\\    
{\large 2. Theoretical Framework}\\
\mbox{ }\\

The starting point is the field equation of general relativity 
\cite{conventions}
\begin{equation}\label{efeq}
G_i^k= \frac{8 \pi G}{c^4}T_i^k
\end{equation}
The energy-momentum tensor is the source of the gravitational field, which in 
turn is embodied in the Einstein tensor $G_i^k$. While the $T_0^0$ component
encompasses all energy density apart from gravity, the latter certainly affects the 
energy content of a system. This is most clearly seen from the mass defect of a 
spherically symmetric ball of matter. We had shown how a neglect of the gravitational contribution can lead to confusion in the tallying of mass
\cite{roscoop}.

Given the fact that gravity affects the energy of a system, we now consider what 
role it may play in the localization of the energy. Various 
authors have presented reasons for regarding the energy, inclusive of the 
contribution from gravity, in a spherically-
symmetric system to be within the region of $T_i^k$ \cite{energyauth} although 
some have argued that it is actually distributed in the entire field 
\cite{alternative}. If it were to be the latter, then the gravitational field 
would be seen to have an energy distribution very much like that of the 
electromagnetic field. However, we will see below that there are reasons to 
believe otherwise.

It is the conservation laws which lend fundamental significance to the energy 
concept and in covariant form, applicable to general relativity, these are
\begin{equation}\label{eq1}
 T_{i;k}^k = 0
\end{equation}
 It is to be noted that this equation has content only in the region of non-vanishing energy-momentum tensor $T_i^k$: in vacuum, it reduces to the empty 
identity 0 = 0. It is when we shift from the local conservation of (\ref{eq1}) 
to 
a form amenable to global conservation with the introduction of the pseudotensor 
$t_i^k$
\begin{equation}\label{eq2}
\frac{\partial}{\partial x^k}\left(\sqrt{-g}(T_i^k + t_i^k)\right) = 0
\end{equation}
in the form of an ordinary vanishing divergence that the unique features of 
general relativistic energy considerations become manifest. 

Unlike $T_i^k$, the pseudotensor, which is constructed from the first partial derivatives of the metric tensor, can be annihilated at any pre-assigned point 
and 
it is this feature which has led probably most researchers from Einstein to the 
present to deny the possibility of a logical energy localization for general 
relativity. However, some authors have come to the position that energy can be 
localized in the $T_i^k$ regions for spherically symmetric sources and some have 
extended this to more general static or stationary sources 
\cite{energyauth}.  

    Many papers have been written which emphasize the similarities between 
electromagnetism and linearized general relativity. However, there are important 
differences which tend to get overlooked. For example, plane gravitational waves 
are seen to be transverse in essence as are plane electromagnetic waves and the 
energy- and momentum-carrying property as seen by the pseudotensor in the 
gravitational case is held in analogy with the energy momentum tensor in the 
electromagnetic case. However, whereas the latter has a generally covariant 
character, the former can not only be annihilated at any pre-determined point, 
as is well-known but also, as is less well-known, it can be expressed in Kerr-
Schild form in which all 
components of the pseudotensor vanish globally. Thus, while the energy and 
momentum of electromagnetic waves is indisputable both theoretically and from 
solid experimental evidence, the situation for gravity waves is clearly not on 
the same footing.

    Another example is that of a spherical ball of charge as compared to a 
spherical ball of mass. In the former case, the energy density as read from 
$T_{00}$ component of the covariant energy-momentum tensor, is distributed 
throughout space whereas in the latter case, there is 
support for the localization of the energy of the system within the region of 
$T_{ik}$, in this case, within the ball itself.

   These considerations led the author to the localization hypothesis
\cite{coop}. An immediate consequence is that gravitational waves would not be 
carriers of energy in vacuum. This is in contradiction to many previous 
calculations which have attributed an energy loss to systems which emit 
gravitational waves and the generally prevailing belief. 

Einstein \cite{einst}(see also \cite{ll}) used the pseudotensor in the context 
of linearized general relativity for weak gravitational fields 
\begin{equation}\label{eq4}
g_{ik}   =  \eta _{ik} + h_{ik}   
\end{equation}
($\eta_{ik}$ is the Minkowski metric and $h_{ik}$ is a small perturbation) to 
demonstrate that a system with a time-varying mass-quadrupole tensor $D_{\alpha 
\beta}$ would lose energy at the rate given by the ``quadrupole formula"
\begin{equation}\label{eq5}
      \dot{E} = -\frac{G}{45c^5}\left( \frac{d^3}{dt^3}
\right) D_{\alpha \beta} \left( \frac{d^3}{d 
t^3}\right) D^{\alpha \beta}
\end{equation}    
This was almost universally accepted and employed through the years, 
particularly in conjunction with systems driven dynamically by non-gravitational 
forces. Challenges were essentially confined to systems such as binary stars 
which are driven by gravity itself \cite{boncoop}.

After Einstein's calculation, Eddington confirmed the result for the specific 
system of a spinning rod \cite{edd} by a direct calculation with the fields and 
the pseudotensor. Later, he performed the analogue of an electromagnetic 
radiation damping calculation for the spinning rod and confirmed both his 
previous result and the quadrupole formula (\ref{eq5}). This will be the focus 
of the following section.\\
\mbox{ }\\

{\large  3. Eddington's Spinning Rod} \\
\mbox{ }\\

Einstein used the Gauss theorem in conjunction with (\ref{eq2}) to express the 
the expected rate of change of the energy in a region in terms of the flux of 
the pseudotensor over the infinite sphere
\begin{equation}\label{eq3}
\frac{\partial}{\partial t}\int \left(\sqrt{-g}(T_0^0 + t_0^0)\right)
 dV = -c \oint\left(\sqrt{-g}t_0^\alpha \right) dS_\alpha
 \end{equation}
This is similar to the familiar procedure in electromagnetism where the equation 
for the radiated energy is
\begin{equation}\label{eq3a}
\frac{\partial}{\partial t}\int \left((T_0^0\right)
 dV = -c \oint\left(T_0^\alpha \right) dS_\alpha
 \end{equation}
However, there are important differences: in (\ref{eq3a}), there is no 
pseudotensor and the flux arises from the \textit{tensorial} 
Poynting vector.

Einstein used (\ref{eq3}) to deduce the quadrupole formula (\ref{eq5})
for the presumed energy loss by a system with a time-varying mass-quadrupole 
tensor in analogy with electric quadrupole radiation arising via (\ref{eq3a}) in 
Maxwell theory.  
Eddington chose to focus upon a specific source, a uniform rod of mass m, length 
2a, spinning with angular frequency $\omega$ in the x-y plane. He solved for the 
gravitational field perturbation $h_{ik}$ and used (\ref{eq3}) to compute an 
energy flux
\begin{equation}\label{eq10}
\frac{dE}{dt} = -\frac{32GI^2\omega^6}{5c^5}
\end{equation}
This result agrees with that computed directly from (\ref{eq5}). Later, in the 
second edition of his book \cite{edd}, Eddington returned to the spinning rod 
and performed the analogue of an electromagnetic radiation damping calculation
using an alternative form of the covariant conservation laws (\ref{eq1}) 
(${\T}^{ab}$ is defined as $\sqrt{-g}T$) 
\begin{equation} \label{eq6}
\frac{\partial}{\partial x^k} \left( {\T}_i^k \right) = \frac{1}{2}
{\T}^{ab} \frac{\partial h_{ab}}{\partial x^i}
\end{equation}
where (\ref{eq4}) was used. For the energy calculation, i is set to 0
and the equation is integrated over the region just beyond the confines of the the source
\begin{equation} \label{eq7}
\frac{\partial}{\partial t}\int{\T}_0^0 dV = \frac {1}{2} \int {\T}^{ab} \frac{\partial}{\partial t} h_{ab} dV
\end{equation}
(note that the 3-divergence contribution from the left hand side of 
(\ref{eq6}) gives no contribution upon integration because of Gauss'
theorem and the choice of integration volume).

 Unlike the 
previous calculations, this approach has the advantage of avoiding the 
pseudotensor entirely and there are no asymptotic conditions to confront.
Moreover, it mirrors the familiar radiation damping calculations of electromagnetism which had proved so successful in unifying the understanding of the role of energy in electromagnetic wave analysis. 
Using the  linearized  form of the Einstein equations (\ref{efeq}) for weak gravitational 
fields in the harmonic gauge, Eddington 
expressed the  retarded  integral 
solution
\begin{equation}\label{eq8}
h_{ab} = -4\int \left[ \frac{T'_{ab}-\frac{1}{2}\eta_{ab}T'}{\left(
r(1-\frac{v_r}{c})\right) }  \right]_{ ret}dV'
\end{equation}
(henceforth, square brackets will indicate retardation by $t-\frac{r}{c}$)
in the manner of  Lienard- Wiechert, which he substituted into (\ref{eq7}). He noted that the resulting integral, now in terms of $T^{ab}$ and $T'_{ab}$,
``exhibits the loss of energy as arising from the mutual action of pairs of elements of the rod, $dV$ and $dV' $". To bring all elements of the resulting expression to a common time t, he performed a 
present-time expansion
\begin{equation}\label{eq9}
\left[ \frac{T'_{ab}}{r\left( 1-\frac{v_r}{c}\right) }
\right]_{ ret} = \frac{T_{ab}'}{r}-\frac{d}{dt}T_{ab}' 
+ \sum_{n=2}^{\infty}\frac{(-1)^n}{n!}\frac{d^n}{dt^n}
\left(r^{n-1}T'_{ab}\right).
\end{equation}
of the retarded part.
Using equations (\ref{eq9}) and (\ref{eq8}) in (\ref{eq7}), Eddington had a
workable form with which to compute the time rate of change of the 
${\T}_0^0$ 
integral. For Eddington, this was the energy loss. We will have more to say about this later. For convenience, he chose to evaluate it at t=0 with the rod,which is spinning in the x-y, plane being along the $x$ axis at t=0. With $dV$ at $x$ and $dV'$ at $x'$, the distance between elements is 
\begin{equation}\label{eq7a}
r=\sqrt{x^2 +{x'}^2 -2xx' cos{\omega t}}
\end{equation}
This is the r to be used in (\ref{eq9})and t set to 0 after the differentiations. Simplifications occur because if $T^{ab}$ is any non-vanishing component at t = 0, ${T'}_{ab}$ will be an even function of time so odd order derivatives vanish. Therefore, he only had to deal with the series
\begin{eqnarray}\label{eq7b}
\frac{\partial}{\partial t}[\frac{T'_{ab}}{r(1-\frac{v_r}{c}}]&=&-\frac{d^2}{dt^2}{T'}_{ab}-\frac{1}{6}\frac{d^4}{dt^4}\left(T'_{ab}(x^2 +{x'}^2- 2xx' cos \omega t)\right)\nonumber \\
&-&\frac{1}{120}\frac{d^6}{dt^6}\left(T'_{ab}{(x^2+{x'}^2-2xx' cos{\omega t})}^2\right)+...
\end{eqnarray}
which is used with (\ref{eq8}) in (\ref{eq7}) and the leading terms
retained. Further simplifications are realized because the rod is symmetrical about the origin and hence in the integration, all terms which are of odd power in either x or x' will not contribute.
He found the 
contributions to (\ref{eq7}) from, respectively, the stress components
$T^{11}$, $T^{22}$, the momentum components $T^{20}$, $T^{02}$ and the
``energy components" $T^{00}$, $T$, the last being the trace $T^k_k$ of the energy-
momentum tensor, which Eddington referred to as the ``proper-density". He 
calculated with $T$ as if it were $T^{00}$ with the rationale that they ``are 
practically the same". However, this is unjustified because the two 
quantities 
differ by the trace of the stress terms ($T_1^1 + T_2^2$) (note that 
$T_3^3$ is zero for 
spin in the x-y plane) and Eddington had already demonstrated that stress 
component products yield contributions to (\ref{eq7}) of the required 
lowest 
order where they appear directly, i.e., outside of the trace $T$ terms. 
Thus, the 
contributions from the stress terms in the trace, $T$, must be confronted.
However, a simple calculation reveals that both the extra terms which 
arise as 
products of stress components with each other as well as the extra terms 
which 
arise as products of stress components with $T^{00}$ cancel each other. 
Thus 
Eddington was fortunate in his neglect of these terms \cite{bones}. 

Upon gathering the various contributions, Eddington found in now the third 
manner of calculation, the expected result (\ref{eq10}).  Apart from the issue of 
whether or not such calculations would be adequate in the case in which the 
rotation is supported by gravity itself, as in a binary system of masses, as 
opposed to the much stronger cohesive forces in a solid continuous distribution 
of mass such as in the rod, most researchers were content with the description 
of radiated gravitational energy loss described above. Indeed it was Eddington \cite{edd} who first recognized that for gravitationally bound systems, $\frac{Gm}{c^2a}$ is of the order $\frac{v^2}{c^2}$ and the linearized approximation is no longer adequate. However, it was the type of calculation by Eddington described above which provided the bedrock of support and faith in the common wisdom that an accelerating system of masses will, in generality, emit energy via gravitational waves in direct analogy to the situation which prevails in electromagnetism. It was the energy 
localization hypothesis which induced the present author to scrutinize the 
standard view, for if energy could flow through vacuum regions as in the case of 
a spinning rod shedding energy as in (\ref{eq10}), the localization hypothesis would be untenable. 

It is the Tolman integral which changes the standard picture.\\
\mbox{ }\\

{\large 4. The Tolman Integral}\\
\mbox{ }\\

Using the pseudotensor, Tolman \cite{tol} showed that for a static or quasi-
static system, the mass could be expressed as
\begin{equation}\label{eq11}
 E=\int\left({\T}_0^0-{\T}_{\alpha}^{\alpha}\right) dV
\end{equation}    
It is to be noted that the work of Tolman \textit{followed} that of Eddington.
Landau and Lifshitz \cite{ll} derived (\ref{eq11}) without recourse to the 
pseudotensor. The proceeded from the identity, valid for time-independent 
systems
\begin{equation} \label{equa1}
R_0^0 =\frac{1}{\sqrt{-g}}\frac{\partial}{\partial x^\alpha}(\sqrt {-g}g^{i0}\Gamma_{0i}^\alpha
\end{equation}
where $\Gamma^a_{bc}$ denotes the Christoffel symbol. They integrated 
(\ref{equa1}) over 3-space and applied Gauss' theorem to convert to a surface 
integral which was evaluated asymptotically with the approximate metric 
functions
\begin{equation} \label{equa2}
g_{00}=1-\frac{2Gm}{c^2r},  g_{\alpha \beta}= - {\delta}_{\alpha \beta}-
\frac{2Gmn_{\alpha}n_{\beta}}{c^2r},    g_{0 \alpha} = 0
\end{equation}
where $n_{\alpha}$ is the unit normal. Evaluating the surface integral yields
\begin{equation} \label{equa3}
\int R_0^0\sqrt{-g}dV= \frac{4 \pi Gm}{c^2}
\end{equation}
Finally, from the field equations (\ref{efeq}), the Ricci tensor component 
$R_0^0$ can be expressed as 
\begin{equation}\label{equa4}
R_0^0=\frac{8 \pi G}{c^4}(T_0^0 - \frac{1}{2}T)= \frac{4 \pi G}{c^4}(T_0^0-
T_{\alpha}^{\alpha})
\end{equation}
and (\ref{eq11}) follows.
There are two points to note: firstly, Tolman succeeded in expressing the total 
energy in the stationary or quasi-stationary state as an integral which is 
confined to the region of $T_{ik}$. This has a particular attraction for us in 
light of the localization hypothesis. Moreover, there are various grounds to 
support such a localization for the stationary case \cite{energyauth}.
Secondly, it is seen that the stresses play a role in the total mass function. 
In weak sources, they would appear to play a negligible role relative to $T_0^0$ 
and in any event, they appear here in conjunction with a mass measure apparently applicable 
to at most quasi-stationary sources. Hence one might conclude that they are doubly 
irrelevant for weak-field radiative energy calculations of the form studied by  
Eddington. We would conjecture that if ever the thought of a role for the stress 
terms might have been entertained in the past in this connection, it would have 
been summarily dismissed for these reasons. However, \textit{it is not the 
magnitude of the stress terms relative to the $T_0^0$ term which is relevant 
here but rather its time rate of change}. Also, while it is true that the Tolman 
integral clearly measures the mass in stationary or quasi-stationary systems, we 
will see how it can be used to extract information in dynamic systems as well.\\ 
\mbox{ }\\

{\large 5. Mass Loss in Dynamic Systems}\\
\mbox{ }\\

While the Tolman integral measures mass in non-dynamic systems, at issue was the 
problem of finding a mass measure for dynamic systems. Bondi \cite{bondi} 
derived a ``news function" to this end but Madore and we  
\cite{coophob} showed that this was actually equivalent to the pseudotensor and 
hence shares in its maladies. It is to be noted that Bondi was careful to state that his mass, which is the mean value of his ``mass aspect" function is \textit{defined} as the mass \cite{bondi}. That it is a definition is frequently overlooked. In any event, Bondi as well as Bonnor 
\cite{bonnor} and Feynman (in a private communication) shared the view that 
the ideal situation would be the following: monitor a system from an initial 
stationary state through a dynamical phase and ending in a final stationary 
state. The mass difference between the initial and final states would reveal the 
unambiguous mass loss. This approach was advocated precisely because one was confronting an essentially different non-tensorial structure in general relativity and doubts had certainly been raised through the years regarding the very foundations of the standard energy scenario. As well, there certainly was no existing experimentation available to support the theoretical deductions.

A more direct approach is available from the vantage point of one who is leaning 
towards the localization hypothesis. Since the Tolman integral measures the mass 
without ambiguity in the initial and final stationary states, the Tolman 
integral must change in the intermediate dynamic phase if there is to be a mass 
loss. This is a \textit{necessary} condition
\cite{necessary}. Thus, the test is one of measuring the change in the Tolman integral in 
the dynamic phase to determine whether the necessary condition for mass loss is met. We do so for the Eddington rod.

Eddington had already computed the ${\T}_0^0$ part of the Tolman integral.
To evaluate the trace of the stresses portion, we will use a variation of the 
method of Einstein \cite{einst}. A more elaborate sequence of steps than that of Einstein is now required because of the demand for greater accuracy in the present context. To this end,
we first multiply (\ref {eq6}) by $g^{il}$, re-express the derivative as
a product and apply the raising operation to get \cite{raising}
\begin{equation}\label{eq16}
{\T}^{lk}_{,k}= F^l
\end{equation}

where

\begin{equation}\label{eq17}
F^l=\frac{1}{2}{\T}^{ab}h_{ab,i}g^{il} +g^{il}_{,k}{\T}_i^k
\end{equation}

We now set $l= \gamma$ and multiply by $x^\delta$ 

\begin{equation}\label{eq18}
x^\delta {\T}^{\gamma 0}_{,0} = -({\T}^{\gamma \beta} x^\delta)_{,\beta}+{\T}^{\gamma \delta} + F^\gamma x^\delta
\end{equation}

Interchanging $\gamma$ and $\delta$, adding the equations, integrating over the source region and applying Gauss' theorem to the 3-divergence terms yields

\begin{eqnarray}\label{eq19}
\frac{\partial}{\partial x^0}\int ({\T}^{\delta 0}x^\gamma+{\T}^{\gamma 0}x^\delta )dV
&=&2\int{\T}^{\gamma \delta} dV+\int({\T}^{\gamma 0}v^{\delta}+{\T}^{\delta 0}v^\gamma)dV \nonumber \\
&+&\int(F^\delta x^\gamma+F^\gamma x^\delta)dV
\end{eqnarray}

where $v^\delta$ is $\frac{dx^\delta}{dt}$

Setting $l=0$ in (\ref{eq16}), multiplying by $x^\gamma x^\delta$, integrating and again applying Gauss' theorem to the 3-divergence term yields

\begin{eqnarray}\label{eq20}
\frac{\partial}{\partial x^0}\int {\T}^{00}x^\gamma x^\delta dV
&=& \int({\T}^{0 \delta}x^\gamma+{\T}^{0 \gamma} x^\delta)dV \nonumber \\
&+& \int {\T}(x^\gamma v^\delta +x^\delta v^\gamma)dV +\int F^0 x^\gamma x^\delta dV
\end{eqnarray} 

After taking $\frac{\partial}{\partial x^0}$ of (\ref{eq20}), eliminating
$\frac{\partial}{\partial x^0}\int ({\T}^{\gamma 0} x^\delta +{\T}^{\delta 0} x^\gamma ) dV$ by using (\ref{eq19}) and setting $\delta$ = $\gamma$ yields

\begin{eqnarray}\label{eq21}
\int {\T}^{\gamma \gamma} dV
&=& \frac{1}{2}\frac{\partial^2}{\partial (x^0)^2}\int {\T}^{00}x^\gamma x^\gamma dV-\int F^\gamma x^\gamma dV -\frac{1}{2}\frac{\partial}{\partial x^0}\int F^0 x^\gamma x^\gamma dV\nonumber \\
&-& \frac{\partial}{\partial x^0}\int{\T}^{00}x^\gamma v^\gamma dV-\int{\T}^{\gamma 0}v^\gamma dV
\end{eqnarray}

To complete the energy loss calculation, we require the time rate of change of the integrated trace of the stresses. This is readily connected to (\ref{eq21}) as follows:

\begin{eqnarray}\label{eq22}
\frac{\partial}{\partial x^0}\int{\T}_\gamma ^\gamma dV = \frac{\partial}{\partial x^0}\int{\T}^{\gamma k} g_{k \gamma}dV
&=&\int {T}^{\gamma \beta}_{,0}g_{\beta \gamma}dV + \int{\T}^{\gamma 0}_{,0}g_{0 \gamma}dV+\int{\T}^{\gamma k}g_{k \gamma ,0}dV\nonumber \\
&=&-\frac{\partial}{\partial x^0}\int{\T}^{\gamma \gamma}dV+\int{\T}^{\gamma \beta}_{,0}h_{\beta \gamma}dV+\int{\T}^{\gamma 0}_{,0}h_{0 \gamma}dV\nonumber \\
&+&\int{\T}^{\gamma 0}h_{0 \gamma,0}dV +\int{\T}^{\gamma \beta}h_{\beta \gamma,0}dV
\end{eqnarray}

where (\ref{eq4}) and $\eta_{\alpha \beta}$= diagonal$(-1,-1,-1)$ have been used.

Finally, with (\ref{eq21}) substituted into (\ref{eq22}) and simplified,
we get the final required expression
\begin{eqnarray}\label{eq23}
\frac{\partial}{\partial x^0}\int {\T}_\gamma^\gamma dV&=& -\frac{1}{2}\frac{\partial^3}{\partial(x^0)^3}\int{\T}^{00}x^\alpha x^\alpha dV+\frac{\partial}{\partial x^0}\int F^\alpha x^\alpha dV\nonumber \\
&+&\frac{1}{2}\frac{\partial^2}{\partial(x^0)^2}\int F^0 x^\alpha x^\alpha dV+\frac{\partial^2}{\partial(x^0)^2}\int{\T}^{00}x^\alpha v^\alpha dV\nonumber \\
&+&\frac{\partial}{\partial x^0}\int{\T}^{\alpha 0} v^\alpha dV+\int{\T}^{\alpha k}_{,0}h_{\alpha k} dV + \int{\T}^{\alpha k}h_{k \alpha ,0} dV
\end{eqnarray}

where $F^l$ is given by(\ref{eq17}) and 
\begin{equation}\label{eq24}
g^{il}=\eta^{il}-h^{il}
\end{equation}
will be used in conjunction with (\ref{eq23}) and(\ref{eq17}).

The complete energy-momentum tensor to lowest order for the spinning rod is required. At $t=0$ when the rod is aligned along the x axis, Eddington \cite{edd} has supplied all but the $T_{11}$ stress component. This is
derived from elementary dynamics with the boundary condition that the tension vanish at the extremities,
\begin{equation}\label{eq25}
T_{11}=\frac{\sigma \omega^2}{2}(x^2-a^2)
\end{equation}
where $\sigma$ is the line mass density $\frac{m}{2a}$. 
From a rotation given by the orthogonal transformation,
\begin{eqnarray}\label{eq26}
x'&=&x cos\omega t -y \sin \omega t\nonumber \\
y'&=&x sin\omega t +y cos\omega t\nonumber \\
\end{eqnarray}
the complete time-dependent $T'^{ik}$ follows:
\begin{eqnarray}\label{eq27}
T'^{11}&=&\frac{\sigma \omega^2}{2}(x^2-a^2){cos^2\omega t} + \sigma \omega^2 x^2 {sin^2\omega t}\nonumber \\
T'^{22}&=&\frac{\sigma \omega^2}{2}(x^2-a^2){sin^2\omega t}+\sigma \omega^2 x^2 {cos^2\omega t}\nonumber \\
T'^{12}&=&-\frac{\sigma \omega^2}{4}(x^2 +a^2)sin 2\omega t\nonumber \\
T'^{01}&=&-\sigma \omega x sin{\omega t}\nonumber \\
T'^{02}&=&\sigma \omega xcos{\omega t}\nonumber \\
T'^{00}&=&\sigma\nonumber \\
\end{eqnarray}

It is to be noted that the $x$ axis is the coincident axis with the rod at
$t=0$ and the $x'$ axis is the coincident axis with the rod at the instant of  time $t$.
Neither axis is co-moving with the rod. Using (\ref{eq27}) and (\ref{eq26}), it is readily demonstrated that the lowest order conservation laws 

\begin{equation}\label{eq28}
T'^{ik}_{,k}=0
\end{equation}
are satisfied, as required. 
The calculation of the Tolman term of (\ref{eq23}) now proceeds using the terms of the energy-momentum tensor of (\ref{eq27}). It is clear that this calculation is considerably more complex than that of Eddington as in the latter, there was only the single integral (which leads to the various factors with summation on a and b) given in (\ref{eq7}). By contrast, we have many integrals to compute for the Tolman contribution as seen in (\ref{eq23}). Moreover, for Eddington, the energy-momentum tensor occurs without a derivative whereas we have to compute terms with derivatives 
of the energy-momentum tensor. It was for this reason that we required more detailed information regarding the energy-momentum tensor than was required by Eddington. However, the basic procedure is the same as that in the Eddington calculation. 

The Gauss theorem is used frequently in the course of the calculations
and time derivatives of integrals are evaluated with the differentiation brought into the integral. This is straightforward if there is continuity but in
the present context, we are dealing with the discontinuous distribution of a line mass. We had dealt with the changes which occur for sources with discontinuities \cite{cooplim} and presented a more detailed version of the quadrupole formula for such sources. Regarding discontinuities in the present problem, we considered the correction to the Gauss theorem

\begin{equation}\label{eq29}
\int dV \nabla .{\bf F}=\oint_S d{\bf S}.{\bf F}+ \oint_D d{\bf S}.{\bf F}|,
\end{equation}
 
the operation of differentiating within an integral of a function f,

\begin{equation}\label{eq30}
 \frac{d}{dt}\int dV f=\int dV f_{,0}+\oint_D d{\bf S}.{\bf v}f|,
\end{equation}

and when further derivatives are called for,
\begin{equation}\label{eq31}
\frac{d}{dt}\oint_D d{\bf S}.{\bf F}|= \oint_D d{\bf S}.\frac{d{\bf F}}{dt}|+\oint_D d{\bf S}.\left({\bf F}(\nabla . {\bf v})-({\bf F}.\nabla){\bf v}\right)
\end{equation}

where $``|"$ denotes inner-minus-outer and ${\bf v}$ is the velocity of the
surface of discontinuity.
 
Very rarely are these refinements considered in the literature. However surely there is a virtue in completeness, particularly in the present case where such a major departure from the prevailing beliefs is being presented. We analyzed the various points in the calculation with the extra contributions arising from the presence of discontinuities and found that they contribute at most higher order corrections. 

Thus, we are left with the calculation of the derivative of the Tolman stress trace integral (\ref{eq23}) using the energy-momentum tensor components (\ref{eq27}), the metric and its derivatives using (\ref{eq8}), (\ref{eq9}), (\ref{eq7a}), (\ref{eq7b}), (\ref{eq24}) and the symmetries discussed previously. Clearly, there are numerous terms to consider, far more than in the case of the Eddington calculation. We had to deal with the gradient of the stresses, which do not enter into the Eddington calculation. Moreover, the trace $T'$ in (\ref{eq8}) as well as the determinant of the metric were taken into account wherever they appeared.  While most of the calculation is routine, there are interesting aspects. For example, in the course of the derivation of the `quadrupole formula' (\cite{einst} \cite{ll}), the derivatives of the spatial components of the energy-momentum components are related to the derivatives of the mass
quadrupole moments using the untraced form of (\ref{eq23}) with only the first term on the right hand side retained. However, for our calculation, we require the 3-trace and since $x^\alpha x^\alpha$ is simply the constant $r^2$ of each element of the rotating rod, we get no contribution from this corresponding term. In passing, we remark that if ever one might be inclined to dismiss
the role of stress terms in the tallying of mass loss in general relativistic calculations, one need only remind one's self as to the origin of the quadrupole formula itself, with the stresses entering as just outlined. Moreover, we are reminded of the importance of the stresses in computing the total energy of a body \cite{roscoop} \cite{stress}.    

For the present calculation, it is from the other terms of (\ref{eq23}) that the non-vanishing contribution is found. After the they are summed, a remarkable result emerges: \textit{ The derivative of the integrated trace of the stresses is precisely the value (\ref{eq10}) that was found by Eddington \cite{edd} 
for the rate of change of the `kinetic energy' ${\T}_0^0$ integral and 
hence the  complete Tolman integral does not change in the dynamic phase to lowest (quadrupole formula) order. The mass is conserved to at least this order.}\\
\mbox{ }\\

{\large 7. Summary and Concluding Discussion }\\
\mbox{ }\\

We began by outlining the unusual role that energy has played in general relativity and the various ideas which have been expressed through the years regarding the issue of its localizability. The reasons which led the present author to hypothesize that energy is most logically localized in regions of non-vanishing energy-momentum tensor $T^{ik}$ were presented. It was noted that if correct, the hypothesis would lead to a hitherto unprecedented aspect of a wave in the case of gravitation: waves carrying real curvature through vacuum would nevertheless be devoid of energy. While such a conclusion might at first glance appear untenable, it was noted that gravity plays a different role in physics from the perspective of general relativity: all particles and fields exist \textit{within} spacetime whereas gravity, in essence, \textit{is} spacetime \cite{coop}.

We discussed the role of the pseudotensor as a vehicle for computing energy loss by a dynamic system and we noted that plane gravitational waves could be expressed in Kerr-Schild form in which all components of the pseudotensor vanish. This was distinguished from plane electromagnetic waves which have tensorial as opposed to pseudotensorial content and hence display an invariantly significant energy flux. This led us to re-consider one of the classical calculations in general relativity, the supposed energy loss from a spinning rod as performed by Eddington. It was noted that Eddington found a mass loss consistent with the quadrupole formula via a flux integral with the pseudotensor as well as with a radiation reaction calculation analogous to that which is performed in electromagnetism. However, it was noted that this was done some years prior to the discovery by Tolman that the trace of the stresses, $T_\alpha^\alpha$, is required in addition to $T_0^0$ to derive the energy of a stationary or quasi-stationary system in general relativity.

Although a spinning rod is not a stationary system, the results of Tolman could be applied as follows: to counter doubts which had been raised through the years and to trace the evolution of a mass loss without ambiguity, experts had argued repeatedly that ideally one should start with a
stationary (or at least quasi-stationary) state for a system which evolves dynamically and returns to an ultimately stationary or quasi-stationary state. The start and finish masses are to be compared and the difference would yield the unambiguous mass loss. Thus we noted that this implies that a
necessary condition for a mass loss is that the Tolman integral
must vary during the dynamic phase, for otherwise the final tally of the Tolman integral would be the same as the start value. We then noted that Eddington had not computed the rate of change of the Tolman integral. Rather, he had computed the rate of change of the integral of ${\T}_0^0$, the `kinetic' energy change. We summarized the elegant manner in which Eddington performed this calculation and we turned our attention to the missing Tolman stress
terms. Unfortunately, the latter is far more difficult to compute than the former and hence we devised a simplifying approach based upon Einstein's classic derivation of the quadrupole formula. In so doing, we found the series of integrals which expresses the rate of change of the Tolman stress-trace integral, which, in conjunction with the Eddington integral, gives the formula for \textit{energy} loss as opposed to \textit{kinetic energy} loss. We found the remarkable result that stress-trace part gives precisely the same result as was found from the Eddington calculation. Since the former appears with a minus sign in the complete Tolman expression, it cancelled the latter. The conclusion was thus that \textit{to the quadrupole formula order, there is no energy loss from a spinning rod, consistent with the localization hypothesis}. With the constancy of the complete Tolman integral, it could then be argued that this integral gives the mass \textit{in generality}, not just in the specialized stationary state. The broader implication of this result was noted, namely that if gravity waves were not energetic, then a re-evaluation of the concept of a graviton, and hence of the quantization of gravity, would be called for.

At this point, it is worth returning to the issue of initial and final states. Since kinetic energy is being lost in the dynamic phase, clearly the process could not have been proceeding from the infinite past. One might posit that the system evolved from a quasi-stationary rotating disk-like configuration with a perturbing element which made it attain a rod shape and hence enter into a dynamic phase. Then the system loses kinetic energy as described above. The previous analysis depended upon the condition $\frac{v^2}{c^2}>>\frac{Gm}{c^2a}$ for its validity and hence a more precise calculation would be required to continue the analysis through the eventual slow-rotation phase. One might envisage two ideal final configurations: a) the rod-like structure reverts to its former disk-like shape through a perturbative element, into a quasi-stationary form or b)it remains rod-like. In either case, there is the issue of angular momentum to consider. We would conjecture that just as energy is conserved, so too will angular momentum, localized within the material distribution, be conserved. A verification to this effect will be the subject of future research. Assuming that this is the case, we recall how a conserved angular momentum can be compatible with a reduced spin rate. What comes to mind immediately is the familiar elementary physics demonstration in which a person holding bar-bells near his body is set into rotation on a bearing-suspended platform. The spin is reduced when the person moves the bar-bells away from his body, increasing the moment of inertia and thus conserving angular momentum. In like manner, we conjecture that in either scenario, a) or b), or whatever other less ideal case, the system will expand to compensate for the slow-down to conserve angular momentum. Since a truly rigid body is incompatible with relativity, this does not present a problem in principle. While in the bar-bell example, kinetic energy is also conserved, in the present rotating rod analysis, the calculations indicate that the kinetic energy is diminished while the stress energy increases. The most reasonable prediction would be that the system asymptotically approaches a quasi-stationary configuration. It would be very interesting if an analysis through the slow spin rate phase could be performed to determine the nature of the final state.

In the bar-bell analogy, kinetic energy as well as angular momentum is conserved whereas the rod loses kinetic energy.
A closer mechanical analogy for our system which has an extra degree of complexity relative to that of the bar-bell example is as follows: consider two masses attached to a spring and encased in a smooth slot holder with traps to keep the masses from moving outwards. The system is set into spinning motion and the traps are programmed to fall away. Then the masses move outwards and the angular velocity and kinetic energy diminish while the stress energy increases with the increasing tension in the spring. The spring stretches out to its ultimate equilibrium length. Angular momentum and total energy are preserved in the process.

To trace the evolution of the Eddington rod through the later phases would appear to be a very interesting (albeit potentially formidable) challenge for future research. 
\mbox{ }\\

 Acknowledgments: This work was supported in part by a grant from the Natural Sciences and Engineering Research Council of Canada. Gratitude is expressed to W.B. Bonnor, A. Chamorro, P. Havas, D.N. Vollick and various colleagues for helpful discussions.

Dedication:   This work is dedicated to the memory of Nathan Rosen.
         
{\small 

\begin{thebibliography}{99}
\bibitem{necessary}
In the strictest sense, one might argue that this does not provide adequate grounds to refer to this as a necesssary condition for mass loss as one might imagine the stress-trace part of the Tolman integral first increases, as we have shown and later, reverses its course and either stays constant or diminishes at a later stage to yield a net mass loss. While there would not appear to be any basis for expecting such a qualitative shift in behaviour, it cannot be ruled out completely in the absence of a calculation to complete the cycle.
\bibitem{conventions}
Latin indices range 0-3, Greek indices range 1-3, G is the gravitational constant, c is the speed of light
\bibitem{stress}
Note that for stationary systems, the energy of a system can be calculated from the volume integral in (\ref{eq3}) and by comparison with
(\ref{eq11}), the Tolman stress integral can be viewed as embodying the essence of the gravitational contribution to energy. To ignore its role in the context of \textit{gravitational} energy loss computations is particularly illogical.
\bibitem{bones}
Many years ago, the late J.L. Synge related to the author how Eddington responded to criticism of some of his mathematical techniques: ``I don't care what they say- I can feel it in my bones!" One can only marvel at the power of Eddington's skeletal receptors.
\bibitem{raising}
It should be noted that the raising and subsequent lowering of indices in the following sequence is required, firstly because the substitution proceeds with raised indices whereas the initial equation has mixed indices and secondly, because the final Tolman expression demands mixed indices so we must revert in the end. In Einstein's derivation, this was trivial because at the lower order, the raising and lowering was adequately performed with the Minkowski metric.
\end{thebibliography}
\end{document}